\documentclass[12pt]{article}
\usepackage{graphics, color}
\usepackage{graphicx}
\usepackage{amssymb, amsmath, tensor}
\pdfoutput=1

\newcommand{\sect}[1]{\section{#1}\setcounter{equation}{0}}

\def\gsim{\, \rlap{$>$}{\lower 1.1ex\hbox{$\sim$}}\,}
\def\lsim{\, \rlap{$<$}{\lower 1.1ex\hbox{$\sim$}}\,}

\makeatletter
\renewcommand\d[1]{\mspace{2mu}\mathrm{d}#1\@ifnextchar\d{\mspace{-3mu}}{}\mspace{4mu}}
\makeatother
\newcommand{\N}{\mathbb{N}}

\begin{document}


\begin{titlepage}
\hfill{NSF-KITP-12-081}
\bigskip
\bigskip\bigskip\bigskip\bigskip\bigskip

\centerline{\Large Bulk and Transhorizon Measurements in AdS/CFT}
\bigskip
\bigskip\bigskip\bigskip

 \centerline{{\bf Idse Heemskerk,$^{*,}$}\footnote{\tt idse@physics.ucsb.edu}
  {\bf Donald Marolf,$^{*,\dagger,}$}\footnote{\tt marolf@physics.ucsb.edu}
   {\bf Joseph Polchinski,$^{*,\dagger,}$}\footnote{\tt joep@kitp.ucsb.edu}
   and 
   {\bf James Sully$\,^{*,}$}\footnote{\tt joep@kitp.ucsb.edu}}
   
\bigskip
\medskip
\centerline{$^*$\em Department of Physics}
\centerline{\em University of California}
\centerline{\em Santa Barbara, CA 93106}
\bigskip
\medskip
\centerline{$^\dagger$\em Kavli Institute for Theoretical Physics}
\centerline{\em University of California}
\centerline{\em Santa Barbara, CA 93106-4030}\bigskip
\bigskip\bigskip


\begin{abstract}
We discuss the construction of bulk operators in asymptotically AdS spacetimes, including the interiors of AdS black holes.  We use this to address the question ``If Schrodinger's cat were behind the horizon of an AdS black hole, could we determine its state by a measurement in the dual CFT?''

\end{abstract}
\end{titlepage}

\baselineskip = 16pt

\tableofcontents

\setcounter{footnote}{0} 
\section{Introduction}

What is the proper framework for quantum gravity, in general spacetimes?  Much of the history of quantum gravity has been a struggle between spacetime locality and quantum mechanics, and thus far locality has gotten the worst of things.  The holographic principle~\cite{'tHooft:1993gx,Susskind:1994vu} suggests that the fundamental entities in the theory should be nonlocal in a radical way.  This is realized in nonperturbative constructions via gauge/gravity duality,\footnote{We include here both Matrix theory~\cite{Banks:1996vh} and AdS/CFT duality~\cite{Maldacena:1997re}, but we will focus on the latter as the duality dictionary takes a more convenient form.}  in which the well-defined dual variables inhabit an ordinary quantum mechanical framework but are highly nonlocal from the bulk gravitational point of view.  Gauge/gravity duality presently describes only spacetimes with special boundary conditions, and the duality dictionary describes in direct way only observations made at the boundary.  It is important to understand its lessons for more general observations and more general spacetimes.

``If Schrodinger's cat were behind the horizon of an AdS black hole, could we determine its state by a measurement in the dual CFT?''  This is a sharp question which helps to illuminate the content of AdS/CFT duality, the meaning of black hole complementarity~\cite{Susskind:1993if,Stephens:1993an}, and the correct framework for quantum gravity in the presence of horizons.  In this paper we will address this and related issues.  Many of our observations have been made previously, but we believe that it is useful to assess what is understood, and what is still missing.

In \S2 we revisit the construction of bulk field operators in terms of operators in the {\mbox CFT}.  We review and elaborate the Green's function method of Hamilton, Kabat, Lifshytz, and Lowe~\cite{Hamilton:2005ju}, applied to AdS spacetimes.  In \S3 we extend this to AdS black hole spacetimes, and address the cat question.  In \S4 we discuss various general issues in reconstructing the bulk, and compare alternate approaches.

Below we assume that the bulk physics is well-approximated by the usual semi-classical equations of motion.  As noted in \cite{Almheiri:2012rt}, this may not be the case for sufficiently old black holes.  We therefore assume that our black holes are young in the sense of \cite{Almheiri:2012rt}.

\sect{Bulk fields in AdS}
\label{map}

The observations of a low energy observer in the bulk can be described in terms of effective bulk field operators.  The one-to-one mapping between bulk and CFT states implies that these operators have images in the \mbox{CFT}.  To construct these, begin with the AdS/CFT dictionary~\cite{GKP,W} in the `extrapolated' form~\cite{BDHM,Balasubramanian:1998de,Harlow:2011ke},
\begin{equation}
\lim_{\rho \to 0} \rho^{-\Delta} \phi(\rho,x) = {\cal O}(x) \,. \label{extrap}
\end{equation}
We will work in global Lorentzian AdS, with coordinates
\begin{equation}
ds^2 = \frac{R^2}{\sin^2 \rho} (-d\tau^2 + d\rho^2 + \cos^2 \rho \,d\Omega^2_{d-1})\,.
\end{equation}
Also, we abbreviate $(\rho,\tau,\vec \Omega) \to (z,x) \to y$.
There is no source here: the $ \rho^{\Delta - d}$ mode of $\phi$ vanishes.  This expresses the local operators of the CFT as the boundary limit of the bulk field operators.  We wish to invert this, with the aid of the bulk field equations.

This is not a standard problem\footnote{Though it is related to certain consequences of Holmgren's uniqueness theorem \cite{Hor}.}: the boundary is not a Cauchy surface for the bulk, and in a sense we are trying to integrate the field equation in a spacelike direction.  Nevertheless the solution exists, at least as a power series in $1/N$, but its form is far from unique.  This is easy to see in the leading planar limit, corresponding to free fields in the bulk, by expanding both sides in Fourier modes~\cite{BDHM,Balasubramanian:1998de,Bena99}.  This approach is less transparent at higher orders in $1/N$, which led to difficulties in these early papers.

This has recently been revisited by Kabat, Lifschytz, and Lowe~\cite{Kabat:2011rz}, who show that there is no obstruction to adding bulk interactions.  We review and elaborate their construction.  Consider first a free bulk field of mass-squared $m^2 = \Delta(\Delta-d)$ where $d$ is the spacetime dimension of the CFT.   Let $G(y|y')$ be any  chosen bulk Green's function,
\begin{equation}
(\Box' - m^2) G(y|y') = \frac{1}{\sqrt{-g}} \delta^{d+1}(y-y') \,.
\end{equation}
Then
\begin{eqnarray}
\phi(y) &=& \int d^{d+1}y'\, \sqrt{-g'} \phi(y') (\Box' - m^2) G(y|y')  \nonumber\\
&=& \lim_{\epsilon \to 0} \int_{\rho' = \epsilon} d^dx' \, \sqrt{-g'} \bigl( G(y|y') \partial^{\rho'} \phi(y') - \phi(y') \partial^{\rho'} G(y|y') \bigr)
+ \mbox{in/out-going} \,. \label{Greensthm}
\end{eqnarray}
We will always use Green's functions whose support is limited to a finite range in global time, so the last term from $\tau = \pm\infty$ is absent, but the reader should note that this might otherwise appear.\footnote{We note that the usual bulk to bulk propagator is not in this class: it does not contain the non-normalizable $K$-mode of (\ref{K}) so that the only contributions in (\ref{Greensthm})  come from the timelike infinities.  }
Near the boundary $\rho'=0$, any Green's function will behave as
\begin{equation}
\label{K}
G(y|y') \to c_\Delta\left( \rho'^\Delta L(y|x') + \rho'^{d - \Delta} K(y|x') \right)\,,
\end{equation}
where we introduce $c_\Delta = R^{1-d}/ (2\Delta - d)$.  Then
\begin{equation}
\phi(y) =  \int d^dx'\, K(y|x'){\cal O}(x')  \,. \label{bulkop}
\end{equation}
If we use another Green's function, differing by a free field solution in $y$, we obtain a different form for the dictionary~(\ref{bulkop}), but the result is necessarily equivalent; we will illustrate this below.

To obtain one useful form for the Green's functions, go to a frame in which $y$ at the center of global AdS, $\rho = \pi/2$.  The Green's function can be taken to be spherically symmetric in $y'$, and so reduces to a function of $\tau'$ and $\rho'$.  Now in this effectively $1+1$ problem we can reverse space and time and integrate radially.  The resulting Green's function is nonvanishing only in a spacelike or null direction from $y$ (Fig.~1a).  For brevity we nevertheless refer to this Green's function as spacelike.
\begin{figure}[!t]
\begin{center}
\vspace {-5pt}
\includegraphics[scale=1.2]{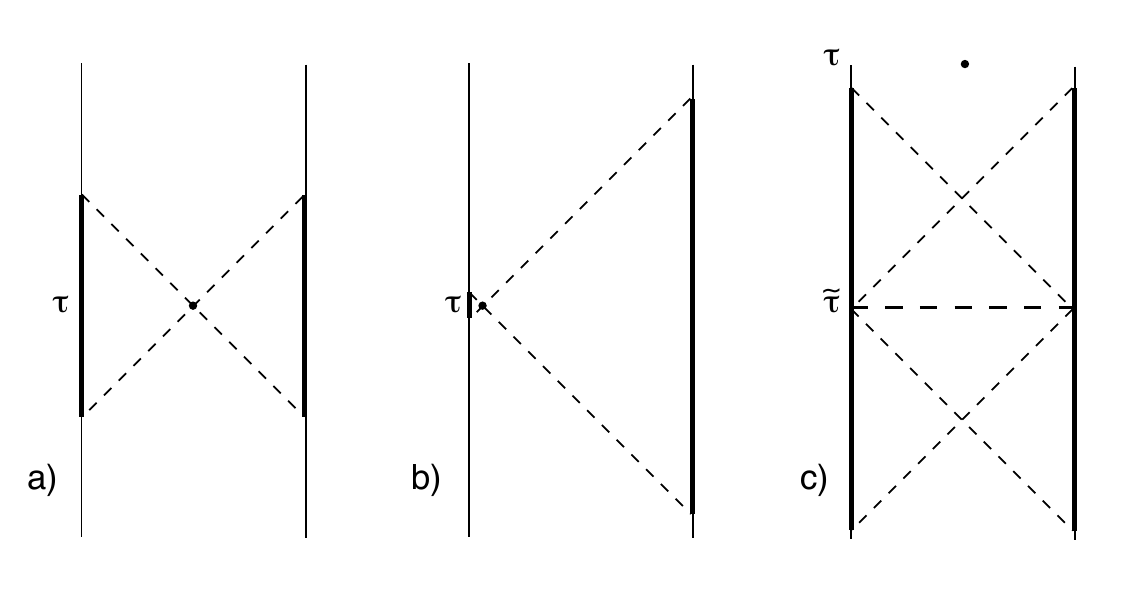}
\end{center}
\vspace {-10pt}
\caption{Boundary constructions of the bulk operator in the center of AdS at time $\tau$, shown as cross sections through global AdS.  The support is indicated in bold. a) An operator in the center of AdS, using the {spacelike} Green's function.  b) An operator elsewhere on the timeslice, obtained by a conformal transformation.  c)
Using ordinary Cauchy evolution from $\tau$ to $\tilde \tau$, and then the {spacelike}  Green's function.
}
\end{figure}
An explicit form for even-dimensional AdS was given in Refs.~\cite{Hamilton:2005ju,Hamilton:2006az}.  In the Appendix we obtain the Green's function for odd-dimensional AdS, and deal with a minor technical issue in the even-dimensional case.

The positive and negative frequency parts of the free field have periodicity $\phi(\rho,\tau + 2\pi,\Omega) = e^{\mp 2\pi i\Delta} \phi(\rho,\tau,\Omega)$, and this is inherited by the operator $\cal O$ in the planar limit.  Using this, we may translate any part of the support of the integral~(\ref{bulkop}) and obtain a different but equivalent form, corresponding to a different choice of Green's function.   For example, we may choose all the support to lie in the range $\tilde\tau-\pi < \tau' < \tilde\tau+\pi$ for some $\tilde \tau$.  Equivalently (Fig.~1c), we may think of this as evolving the bulk field from $\tau$ to $\tilde\tau$ by ordinary Cauchy evolution, and then using the Green's function construction with the spacelike Green's function.\footnote{Using $\phi(\rho,\tau + \pi,\Omega) = e^{\mp \pi i\Delta} \phi(\rho,\tau, -\Omega)$, one could further restrict support to a single Poincar\'e patch.}

It is instructive to extend this to the product of two bulk fields
$ \phi(y_1) \phi(y_2) $,
where we take $\tau_1 > \tau_2$ so this is implicitly time ordered.  We can write this in two different forms
\begin{eqnarray}
\phi(y_1) \phi(y_2)
&=&  \int d^{d+1}y\, \sqrt{-g} \, {\rm T}\!\left( \phi(y) \phi(y_2) \right) (\Box_y - m^2) G(y_1|y)
\label{timeo}\\
&=&  \int d^{d+1}y\, \sqrt{-g} \,\phi(y) \phi(y_2) (\Box_y - m^2) G(y_1|y)  \label{wight}
\end{eqnarray}
These are equivalent because the two products coincide where $(\Box_y - m^2) G$ is nonzero (at $y_1$).
The time-ordered product~(\ref{timeo}) leads to
\begin{eqnarray}
\phi(y_1) \phi(y_2)
&=&  \int d^dx\, K(y|x)\,{\rm T}\left( {\cal O}(x)  \phi(y_2) \right) + \int d^{d+1}y\, \sqrt{-g} G(y_1|y)  (\Box_y - m^2) \,{\rm T}\left(\phi(y) \phi(y_2) \right)
\nonumber\\
&=&\int d^dx\, K(y_1|x) {\rm T}\left( {\cal O}(x)  \phi(y_2) \right) +  i G(y_1|y_2)
\nonumber\\
&=&\int d^dx_1\,d^d x_2\, K(y_1|x_1) K(y_2|x_2)  {\rm T}\left( {\cal O}(x_1) {\cal O}(x_2) \right) +  i G(y_1|y_2)
\,, \label{timefin}
\end{eqnarray}
where we have taken $\phi$ to be canonically normalized.
The Wightman product~(\ref{wight}) leads to
\begin{eqnarray}
\phi(y_1) \phi(y_2)
&=&  \int d^dx\, K(y|x) {\cal O}(x)  \phi(y_2)  + \int d^{d+1}y\, \sqrt{-g} G(y_1|y)  (\Box_y - m^2)  \phi(y) \phi(y_2)
\nonumber\\
&=& \int d^dx\, K(y_1|x)  {\cal O}(x)  \phi(y_2) \nonumber\\
&=& \int d^dx_1\,d^d x_2\, K(y_1|x_1) K(y_2|x_2)  {\cal O}(x_1) {\cal O}(x_2) \,.
\end{eqnarray}
Taking the difference of the two right sides, we also get
\begin{equation}
 i G(y_1|y_2) =  \int d^dx_1\,d^d x_2\, K(y_1|x_1) K(y_2|x_2) \theta(\tau_{x_2} - \tau_{x_1})\left[{\cal O}(x_1), {\cal O}(x_2)\right]
 \,, \label{wightfin}
 \end{equation}
 which is not obvious but must be true.  In particular, the singularity at $y_1 = y_2$ must come from the integral near the light-cone.

The final expressions have potential divergences from coincident points~\cite{FSS}.  The Wightman form~(\ref{wightfin}) makes it clear that these are actually not present.  Deform the $\tau_{x_1}$ contour by $-i\epsilon$:
\begin{equation}
\phi(y_1)
\phi(y_2) = \int d^dx_1\,d^d x_2\, K(y_1|x_1 - i \epsilon \hat\tau) K(y_2|x_2)  {\cal O}(x_1 - i \epsilon \hat\tau) {\cal O}(x_2)
\end{equation}
This is well defined for the Wightman product because it corresponds to inserting the convergence factor $e^{-\epsilon H}$ between the operators.    The coincident points can thus be avoided, and there is no divergence.\footnote{We thank J. Kaplan and A. Katz for discussions.  This argument applies at points where the $K$ functions are smooth.  The collision of operators at singularities of $K$ is need to produce the singularities of the Green's function~(\ref{wightfin}).}

Now consider an interacting field, using the simplest cubic interaction for illustration,
\begin{equation}
(\Box - m^2)\phi = \frac{g}{N}\phi^2\,.
\end{equation}
We have normalized the scalar canonically in order to make the $N$ dependence manifest.
Green's theorem gives
\begin{equation}
\phi(y) =  \int d^dx_1\, K(y|x_1){\cal O}(x_1) + \frac{g}{N} \int d^{d+1}y'\, \sqrt{-g'} G(y|y') \phi^2(y') \,.
\end{equation}
Now iterate in the $\phi^2$ term.  This can be put in two forms, according to whether we use (\ref{timefin}) or (\ref{wightfin}):
\begin{eqnarray}
\phi(y) &=& \int d^dx_1\, K(y|x_1){\cal O}(x_1)  \nonumber\\
&&
+ \frac{g}{N} \!\int\! d^{d+1}y'\,d^dx_1\,d^d x_2\, \sqrt{-g'} G(y|y') K(y'|x_1) K(y'|x_2)  {\rm T}\left( {\cal O}(x_1) {\cal O}(x_2) \right)   \nonumber\\
&&
 + \frac{ig}{N}\! \int\! d^{d+1}y'\, \sqrt{-g'} G(y|y') G(0) + O(1/N^2)
\nonumber\\
&=&\int d^dx_1\, K(y|x_1){\cal O}(x_1) \nonumber\\
&&
 + \frac{g}{N} \!\int\! d^{d+1}y'\,d^dx_1\,d^d x_2\, \sqrt{-g'} G(y|y') K(y'|x_1) K(y'|x_2)  {\cal O}(x_1) {\cal O}(x_2)  + O(1/N^2)\,.
 \qquad\ \  \label{intop}
\end{eqnarray}
Expressed in terms of the Wightman CFT product, only tree graphs appear in the construction.  Expressed in terms of time-ordered CFT products, one sums over loops as well, here the tadpole graph.\footnote
{This is proportional to the divergent $G(0)$, so our effective bulk equations of motion must be supplemented by a renormalization scheme, including field renormalization, matched onto the full string theory.   Incidentally, the tadpole graph also has an IR divergence from the integral over AdS spacetime.  Indeed, if a marginal field like the dilaton were to have a tadpole, the AdS asymptotics would be spoiled.  This problem is avoided in examples where supersymmetry forbids the tadpole, or where the dilaton is stabilized.}
Again, use of different Green's functions gives different but equivalent forms.\footnote
{The interacting fields no longer have a simple periodicity in $\tau$, but there is a nonlinear periodicity relation.  In the CFT, the operators of definite dimension are linear combinations of single- and multi-trace terms.}

Clearly this can be iterated to any order in $1/N$.   Of course, we expect many subtleties nonperturbatively in $N$, and one does not expect to be able to define exact bulk observables.  For the purposes here we will be satisfied with the accuracy of the $1/N$ expansion.  It should be noted also that our whole discussion is framed in the gravity limit, corresponding to strong coupling in the gauge theory.

Starting from a bulk description, where the fields have canonical commutators, the usual AdS/CFT dictionary~(\ref{extrap}) constructs boundary operators in the \mbox{CFT}.  The inverse dictionary~(\ref{intop}) reconstructs the original fields, and so these have canonical commutators, vanishing at spacelike separation (up to gauge subtleties to be discussed shortly).  In Sec.~4 we discuss the possibility of a less circular construction, but for now we recall that symmetry determines the form of the CFT two-and three-point functions completely.  Thus at zeroth~\cite{BDHM,Balasubramanian:1998de,Bena99,Hamilton:2005ju} and first~\cite{Kabat:2011rz} orders in $1/N$ one can recover bulk locality starting from a general \mbox{CFT}.  At the next order bulk locality constrains the form of the CFT correlator~\cite{Polchinski:1999ry,Susskind:1998vk,Gary:2009ae,Heemskerk:2009pn}, and local fields can only be recovered to the extent to which this is satisfied.

The bulk theory is general coordinate invariant, and usually has ordinary gauge invariances as well.  The CFT operators are invariant under these, so the bulk fields must be constructed in a fully gauge-fixed form, i.e. a physical gauge.  The above construction applies to the metric and other nonscalar fields, in any given gauge.  As a result, the commutators cannot be strictly local.  One can readily see how this will come out of the construction, taking the example of a gauge symmetry.  The extrapolate dictionary is
\begin{equation}
\lim_{\rho \to 0} \sqrt{-g} F^{\rho0}(\rho,x) = j^0 \,. \label{gauge}
\end{equation}
Consider any bulk field charged under the symmetry.  The charge $Q = \int d^{d-1} {\vec x}\, j^0$ will have a nonzero commutator with this field.  But then~(\ref{gauge}) implies that charged operators anywhere in the bulk will have nonzero commutators with the electric field near the boundary.  So when we construct local operators, it is understood that their commutators are local only to the extent allowed by Gauss's law.  Recall that this issue is especially important in gravitational theories where, due to the universal coupling to energy, {\it any} operator confined to a finite region of space-time is charged under gravity.   In particular, in the gravitational context there is no analogue of compactly-supported Wilson loop observables that might allow one to avoid this issue.  A more detailed treatment of this issue will appear in Refs.~\cite{Heemskerk:2012mq,Kabat:2012hp}

In order to express the measurement in terms of a single-time Hilbert space, we now need to integrate the CFT operators to some reference time.  For example, we might choose to make the measurement at a time $\tau$ spacelike separated from the bulk operator.  The expression~(\ref{intop}) involves only products of local gauge invariant operators, but time evolution will generate nonlocal gauge invariant operators.  For example, for ${\cal O}(x) = {\rm Tr}(F_{\mu\nu}(x)F^{\mu\nu}(x))$, even in free field theory the two constituent fields will propagate outward independently, while interactions will generate further nonlinearities.  These nonlocal gauge invariant operators are referred to as {\it precursors}~\cite{Polchinski:1999yd}. Presumably  they can be expanded in a basis of Wilson loops, though there are subtleties as we will discuss further in Sec.~4.2.3.

As an aside, it is not guaranteed by holography that  bulk fields can be expressed of in terms of products of {local} gauge invariant operators such as we have found.  The identity of the bulk and boundary Hilbert spaces implies a mapping of operators, but it need not take such a local form, and in other situations it might not.

\sect{Black holes}

We would like to be able to extend the above construction to other classical spacetimes with AdS boundary conditions, including time-dependent ones.  In principle this is already implicit in the expansions~(\ref{timefin}, \ref{wightfin}): these are operator statements, and so the change of background is accounted for by the expectation values of the $\cal O$.  However, in a classical background the expectation values of $\phi$ and $\cal O$ are of order $N$ in our canonical normalization, and so $N$ is not a  small parameter in the expansion.  But for backgrounds sufficiently close to AdS we should still be able to construct $\phi$ from the expansion.

The background we would particularly like to study is a small black hole that forms from diffuse matter and then decays.  For fields behind the horizon, ordinary Cauchy evolution can be used to express them in terms of operators before the black hole formed, when the geometry was close to AdS, and then the above construction can be used (Fig.~2).
\begin{figure}[!t]
\begin{center}
\vspace {-5pt}
\includegraphics[scale=.8]{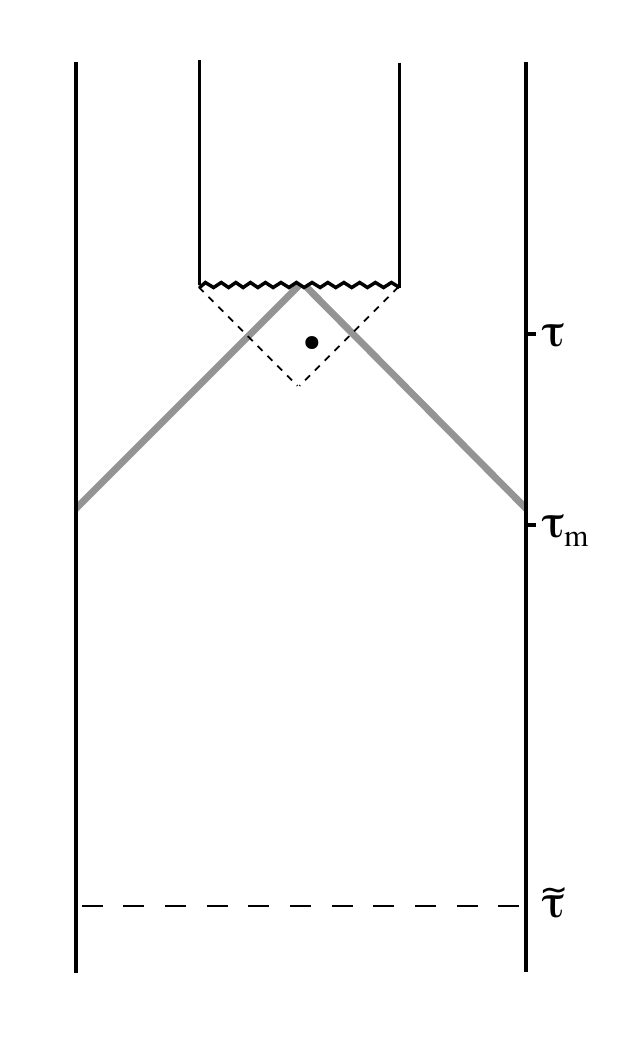}
\vspace {-10pt}
\end{center}
\caption{Formation and evaporation of an AdS black hole.  The grey lines represent an ingoing null shell, formed by perturbing the CFT.  Field operators behind the horizon are integrated backwards in the bulk to before the formation of the black hole, and then expressed in terms of CFT operators; these can be integrated forward e.g.\ to times $\tau$ or $\tau_{\rm m}$.    The Penrose diagram is doubled to match Fig.~1.}
\end{figure}
 Thus we conclude that a relation of the form~(\ref{intop}) holds for field operators behind the horizons of such black holes.

This construction immediately answers the cat question.  Using these field operators we can construct projections onto live and dead cats (or simply measure the cat's temperature!).  A simpler version of this argument was presented in Ref.~\cite{Freivogel:2004rd}.  Imagine throwing a pair of particles into a black hole, such that they collide behind the horizon.  The scattering angle is then the quantum variable, and is readily measured using field operators.  Refs \cite{Marolf:2008mg} described a similar measurement involving a spin.

It should be noted that accuracy of order $1/N$ is sufficient to make the measurement with reasonable confidence: we do not need accuracy $e^{-N}$ in the bulk evolution.  On the other hand, the boundary evolution is assumed to be exact, we are assuming that we can solve the CFT.

A system behind the horizon can be described in a Hilbert space constructed using the local bulk operators.  Through the above dictionary these act in the Hilbert space of the \mbox{CFT}, so the interior Hilbert space is embedded within that of the \mbox{CFT}.  One important moral is that the CFT is dual to the whole of the bulk, it does not really live at the boundary even though its spacetime is isomorphic to the (conformal) boundary.  Of course many authors have made these points from various perspectives, including Refs.~\cite{BDHM, Balasubramanian:1999zv,Susskind:1999ey,Maldacena:2001kr,Hubeny:2002dg,Kraus:2002iv, Fidkowski:2003nf, Freivogel:2004rd, Hamilton:2005ju, Hamilton:2006fh, Marolf:2008mg, Marolf:2008mf, Horowitz:2009wm}.  But it is an important one that deserves to be reexamined and sharpened if possible.

This argument also distinguishes two notions of black hole complementarity.  Quantum gravity as an effective theory describes spatial slices that extend through the horizon of a black hole, intersecting both the outgoing Hawking radiation and the quanta falling toward the singularity.  For example, the wavefunction in canonical gravity would assign amplitudes to such geometries.  Black hole complementarity~\cite{Susskind:1993if,Stephens:1993an} asserts that this is very wrong.   Except perhaps at some coarse-grained level, it is not even an approximation to the actual Hilbert space, in that the
latter must be much smaller than the tensor product of the Hilbert space
behind the horizon and that of the Hawking radiation.

This is a negative statement, but there is a stronger positive one as well: that there is {\it some} Hilbert space, which is fundamental in the formulation of the theory. The interior and exterior Hilbert spaces are both embedded in this, but not as a product, so the interior and exterior operators do not commute.  In this strong form, the framework of quantum mechanics remains fully intact, but locality is badly broken down.  This appears to be the lesson of AdS/CFT: the field operators behind the horizon can be expressed in terms of CFT operators and then evolved forward in time until the black hole has evaporated; thus they act also in the Hilbert space of the Hawking radiation.

The assertion that we can measure events behind the horizon evokes strong reactions, from ``Obviously" to ``Obviously not.''  Let us address some objections.  If we were talking about an event in the center of AdS, we could of course observe from the boundary at later times and see if and when the cat died.  But we can take those same measurement operators and evolve them backwards in the CFT.  As long as we act after whatever operators were used to prepare the cat, for example at a spacelike separation from the event, we are guaranteed to get the same answer from looking after the fact as from `simultaneous' measurement.

The possibility of direct observation from the boundary at later times is not present for the black hole, but the situations are not really so different~\cite{Hubeny:2002dg}.  In Fig.~2, we can measure the marked event behind the horizon at the time $\tau_{\rm m}$, on a common spacelike slice with the event.  Note, however, that the infalling shell that forms the black hole is produced later.  We can choose not to send in the shell, and confirm our measurement by direct observation from the boundary, or send it in and hide the event behind a horizon: we can no longer check the result, but the situation on the spacelike slice of the event and measurement is exactly the same.

Of course there is no unique mapping of the time of a bulk event to the time of a boundary event, nor do we assume one.  The precise time of the boundary measurement is irrelevant, since we can evolve CFT operators in time; only its order with respect to other measurements matters.  In the bulk, the time of the measurement is built into the construction~(\ref{bulkop}) of the dual operator.

For example, consider again a measurement at the marked point inside the horizon in Fig.~2. To be concrete let us imagine we are measuring a spin. Regardless of the bulk coordinates we use to label the event, the construction of the previous section yields a non-local boundary operator at some arbitrary boundary time that measures the spin. Conversely, we can construct operators at that same boundary time that measure  spin at points in the future or past of the marked point. Depending on the bulk coordinate system, one of these points could be labeled by the same time coordinate as the boundary operator but this is irrelevant. At any boundary time there is a family of boundary spin operators labeled by the time they measure in the bulk.\footnote{Nevertheless, as we discuss in section \ref{derive} below, some notions of mapping between CFT and bulk are preferred in the sense that they depend at most locally on the choice of CFT dynamics.}

The boundary operators dual to a bulk field can be expressed as either local operators smeared over a range of positions and times, or nonlocal operators at a single time.  If we consider observers who `live' in the CFT (or some QFT coupled to it), we would assume that they are constrained by causality to make the usual kinds of local observations.  In this case there could be limitations to bulk measurements coming from the time it takes to make the CFT measurements;
see e.g. \cite{Sorkin:1993gg,Beckman:2001ck}.  However, we are asking a different question: we want to know what is encoded in the state of the CFT, independent of any such locality constraint.  Thus we are imagining a meta-observer who is free to couple their measuring apparatus to any gauge-invariant operator, local or not, in the Hilbert space of the CFT at some time.  Indeed, if we had considered Matrix Theory instead of AdS/CFT, the dual theory is just quantum mechanics and there would be no such causality issue.

\section{Deriving the bulk}
\label{derive}

\subsection{Uniqueness of the bulk-boundary map}

The preceding construction is instructive.  However, it does not seem fully satisfactory.  In order to construct the local fields in the bulk, and to relate them to operators in a single-time Hilbert space, we need to be able to solve {\it both} the bulk and boundary dynamics.\footnote{In particular our construction is perturbative in $1/N$.  If there is some unexpectedly large nonperturbative effect in the black hole interior, such as discussed in Ref.~\cite{Almheiri:2012rt} for ``old'' black holes or perhaps in the fuzzball proposal~\cite{Mathur:2009hf}, then the interior constructed using the $1/N$ expansion is fictitious.
}  For example we integrate backwards in the bulk, and then forwards in the CFT, to construct the bulk operator in terms of some single-time Hilbert space.  In AdS/CFT we are accustomed to thinking that an exact solution to the CFT gives us a full construction of the bulk dynamics, but this is only for the boundary limit of the bulk observables.  Thus, for example, if we wished to determine the curvature tensor at some point in the bulk, the construction starts by integrating the bulk field equations to the boundary: we are simply {\it calculating} it from the boundary data.

It would seem preferable to have some intrinsic way to identify the bulk field operators, for example through their property of commutativity at spacelike separation~\cite{Kabat:2011rz} (or more precisely, commutativity up to Gauss's law tails),
\begin{equation}
[\phi(y), \phi(y')] \approx 0 \,, \quad d^2(y,y')> 0 \,.
\end{equation}
By itself, this is not enough: if we have such operators $\phi(x)$ indexed by the points of some spacetime, then
\begin{equation}
\phi'(y) = U^{-1} \phi(y) U \label{commutes}
\end{equation}
have the same property, for any unitary $U$.  But if we supplement this by the dictionary~(\ref{extrap}), $\lim_{\rho \to 0} \rho^{-\Delta} \phi(\rho,x) = {\cal O}(x)$, we have further that $U^{-1} {\cal O}(x) U = {\cal O}(x)$ for every local operator, which implies that $U$ must by the identity.

A more general set of spacelike-commuting operators would be generated as follows.  Begin with a set $\phi(\rho,0,\vec x)$ at $\tau = 0$ (or on any spacelike slice).  Now evolve them forward with some relativistic hamiltonian $H'$, which may differ from the actual Hamiltonian $H$.  The the fields $\phi'$ generated by $H'$ and $\phi$ generated by $H$ are related
\begin{equation}
\phi'(\rho,\tau,\vec x) = U^{-1}(\tau) \phi(\rho,\tau,\vec x) U(\tau) \,,\quad U(\tau) = e^{iH\tau} e^{-i H'\tau} \,.
\end{equation}
Again the fields commute at spacelike separation, by the assumption that $H'$ is relativistic (in some metric).  In order for both sets of fields to lead to the same dual operators ${\cal O}(x)$, $H$ and $H'$ must be chosen so that $U(\tau)$ commutes with these local CFT operators at the same time $\tau.$  This is a weaker condition than the previous~(\ref{commutes}), which required that a fixed $U$ commute with the local CFT operators at all times $\tau$,
and is not enough to conclude algebraically that $\phi' = \phi$.  But let us introduce
the further assumption that $H$ and $H'$ are integrals of local densities constructed from the local fields.  Requiring that the dynamics generated by $H$ or $H'$ is consistent with that generated in the CFT should then be enough to determine it completely, $H' = H$: calculating correlators in the CFT, and matching them to bulk calculations, allows us to fix the parameters in $H$ to any given accuracy.\footnote{This argument has previously been made by L. Susskind (unpublished).}

So we can give a more intrinsic description of the bulk fields: they must commute appropriately at spacelike separation, evolve under a local Hamiltonian, and be consistent with the boundary dictionary~(\ref{extrap}).  The previous sections then give a construction.  The boundary dictionary~(\ref{extrap}) plays an essential role here in determining the bulk operators: we need to start with a large set of known observables, at the boundary, to build upon.  In a more general situation, without a special boundary such as that of AdS, it is not clear what conditions would determine the local bulk fields within the space of dual operators.

It is interesting to consider the bulk region that is spacelike with respect to some particular boundary time slice $\tau_0$, Fig.~3.
\begin{figure}[!t]
\label{diamond}
\begin{center}
\vspace {-5pt}
\includegraphics[scale=1.3]{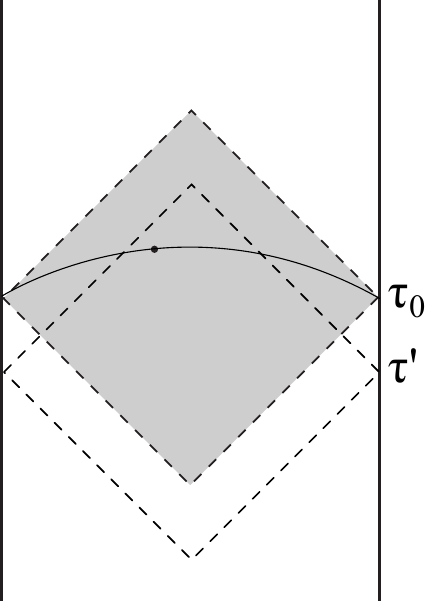}
\end{center}
\vspace {-10pt}
\caption{The mapping from operators in the bulk region that is spacelike with respect to $\tau_0$ to CFT operators at $\tau_0$ is independent of the Hamiltonian at other times. Because the operator at the marked position is really defined non-localy on a Cauchy surface for the diamond, it does not fit into any other diamond.}
\end{figure}
If we wish to represent the operators within this region in terms of CFT operators at $\tau_0$, we have to integrate in the bulk out of this region to the boundary, and then back to $\tau_0$ in the CFT.  The result would seem to depend on the CFT Hamiltonian at other times, and so on measurements that we make before or after by perturbing it.  However, so long as the result is unique when this data is specified, the result cannot actually depend on the given data.  To see this, note that we could find the mapping by first evolving forward in time in the bulk, and then backwards in the CFT: this does not depend on the Hamiltonian before $\tau_0$.  Similarly we can conclude that the result does not depend on the Hamiltonian after $\tau_0$.  It follows that the mapping from bulk to CFT at spacelike-related times does not depend on observations at other times, or on the Hamiltonian at other times.

It is also interesting to examine this point from the perspective of \cite{Marolf:2008mf},  which noted that the construction relating bulk operators to spacelike-separated CFT operators at $\tau_0$ can also be interpreted as relating two operators living in the bulk gravitational theory.  As above, one uses the bulk equations of motion to express one operator (${\cal O}^{grav}_1$) defined inside the diamond of figure \ref{diamond}  in terms of the boundary values of bulk fields at later times.   {\it Bulk}  time translations can then be used express these boundary values in terms of the boundary values of bulk fields at $\tau_0$.  This expression defines the second operator ($O^{grav}_2$) which, because the gravitational Hamiltonian is itself a boundary term, is built only from boundary values of bulk fields at $\tau_0$.   Yet by construction ${\cal O}^{grav}_1 = {\cal O}^{grav}_2$.

To connect this to our discussion above, recall that the only equations of motion which can relate two operators on a single Cauchy surface\footnote{For the diamond.  There are of course no Cauchy surfaces for AdS.} are the canonical constraints.  As a result, while it was derived by evolving operators both forward and backward in time, the relation
${\cal O}^{grav}_1 = {\cal O}^{grav}_2$ must in fact follow from the constraints on a single Cauchy surface\footnote{It would be very interesting to find an explicit construction using only such constraints.}.  Furthermore,
since ${\cal O}^{grav}_2$ is expressed in terms of boundary values of bulk fields at $\tau_0$, it has a clear transcription as a CFT operator at $\tau_0$.  Since the only equations of motion required are constraints, it follows that the map is  independent of the dynamics (either bulk or boundary) at other times.

\subsection{Subtle points of a unique map}

Although this result may seem very natural, it brings to the fore a number of subtleties.  Suppose for example  that a single spin propagates deep in the bulk and that we are interested in its value $\sigma_\rho(\tau_0)$ at some global time $\tau_0$.   To the extent that $\sigma_\rho$
is a local operator in the bulk, it lies in the diamond associated with any boundary time $\tau$ in the range $(\tau_0 - \epsilon, \tau_0 +\epsilon)$ for $\epsilon$ sufficiently small.
But then the arguments above would seem to imply that for any such $\tau, \tau'$ we could relate
$\sigma_\rho(\tau_0)$ to CFT operators $X_\tau,X_{\tau'}$ at $\tau, \tau'$ so as to conclude $X_{\tau} = X_{\tau'}$ without regard to  the CFT dynamics.   But it is easy to modify any such relation by changing the CFT dynamics; e.g., by measuring\footnote{The idea of measuring bulk observables by measuring spacelike related CFT observables leads to many interesting issues.  We refrain from digressing on this point here but instead refer the reader to \cite{Marolf:2008mg}. }
a non-commuting operator such as the one dual to $\sigma_\rho(\tau_0)$.

To resolve this tension we must be more precise in our statements concerning the bulk-boundary map.  The duality is often taken to relate gauge-invariant on-shell bulk observables to gauge-invariant on-shell boundary observables.\footnote{The statement that we consider gauge-invariant operators was implicit in our assumption above that, at least once the CFT dynamics is specified for all time, the boundary image of a bulk operator is unique.} But whether two observables agree on-shell will clearly depend on the choice of CFT dynamics.  To discuss a map that might not depend on this choice we must extend the correspondence to off-shell observables.  While the notion of off-shell gauge-invariant observables is clearly well-defined in the CFT, it is less so on the bulk side of the correspondence.  The issue is most simply stated in the canonical formalism, where the Hamiltonian constraint generates local time translations.  Since gauge invariant observables must commute with this constraint, some part of the bulk dynamics must be specified in order to define them.  In particular, in order to discuss gauge invariant observables we must exclude the possibility that bulk sources might act in the part of the spacetime described by such observables.

Let us therefore recall two facts about the Hamiltonian constraint.  First, in the canonical formalism one considers data on spacelike slices.  Second, only transformations that vanish at the boundary are pure gauge.  We therefore propose that the correct semi-classical notion of a bulk ``off-shell gauge-invariant observable" is one defined by choosing some globally hyperbolic region $R$ of the bulk whose boundary is a Cauchy surface of the boundary spacetime.  Any such $R$ is like a cosmological spacetime in which all local time translatons are pure gauge.  No modifications of the dynamics will be allowed in the interior of $R$, so there is a well-defined notion of bulk gauge-invariance within $R$.  Since $R$ is globally hyperbolic, this notion is independent of any boundary conditions on bulk fields.  In this sense the notion remains ``off-shell," as the bulk dynamics beyond $R$ is determined by the choice of appropriate boundary conditions.   We propose that the correct notion of an off-shell bulk to boundary dictionary relates bulk observables defined within $R$ to CFT observables on the boundary of $R$.

Returning to the example in figure \ref{diamond}, either diamond defines such a globally region $R$.  Let us call the two regions $R_{\tau_0}, R_{\tau'}$.
Noting that the marked point lies in both diamonds, one might be tempted to think that the  ``same'' bulk observable
$\sigma_\rho(\tau_0)$ could be associated with either $R_{\tau_0}$ or $R_{\tau'}$.  This is the source of the confusion mentioned at the beginning of this subsection.  However, we believe that this is not the case, and that there is simply no sense in which an observable associated with $R_{\tau_0}$ can be identified with an observable associated with $R_{\tau'}$ without specifying boundary conditions on bulk fields between $\tau'$ and $\tau_0$.\footnote{
This belief is motivated by concerns associated with dynamical gravity in the bulk (see below).  Note that bulk gravity must be treated dynamically if one is to make use of the CFT stress tensor or Hamiltonian, as we have done in constructing dual CFT observables localized at a specific time $\tau$.  In contrast, 
discussion with the authors of \cite{Bousso:2012sj} suggests that their arguments are restricted to more narrow contexts in which bulk gravity may be treated as non-dynamical.  Thus our construction cannot be performed within their approximation. 
}

At some level this is a proposal.  But it is supported by bulk investigations of locality in quantum gravity
\cite{Giddings:2005id,Giddings:2007nu} which argued that precisely local observables fail to exist in a dramatic sense even at first non-trivial order in the bulk Planck length.\footnote{This effect is inherently field-theoretic.  There are also senses in which locality fails in 0+1 (quantum mechanical) systems with time-reparametrization invariance, though these are somewhat more subtle \cite{Rovelli:1990jm,Rovelli:1990ph,Rovelli:1990pi,Rovelli:2001bz,QORD,MarT}.}  In particular,  since the regions $R_{\tau_0}$, $R_{\tau'}$ may be thought of as non-compact time-dependent cosmologies, attempts to construct local observables in $R_{\tau_0}$, $R_{\tau'}$ are subject to the same infrared divergences noted in \cite{Giddings:2005id,Giddings:2007nu}.   Now, as suggested in \cite{Giddings:2007nu}, one should nevertheless be able to define observables which localize in the limit $\ell_p \rightarrow 0$ by making use of global information associated with an entire bulk Cauchy surface.  But since $R_{\tau_0}$, $R_{\tau'}$ have no Cauchy surfaces in common, this procedure necessarily associates a given bulk observable with at most one of these regions.

\subsection{A path integral perspective}

The path integral gives a useful perspective on the construction of the bulk operators.  Recall that we have first represented these as local operators smeared over some spacetime region of the boundary as in Eq.~(\ref{bulkop}), and then used the boundary evolution to express them in terms of a single-time Hilbert space, 
\begin{equation}
\label{ev}
{\cal O}(\tau, x) = e^{-i H (\tau-\tau_0)} {\cal O}(\tau_0, x) e^{i H (\tau-\tau_0)} .
\end{equation}
 We can think about the last step in two ways.  First, in terms of an explicit CFT dual, {dd the right-hand side of \ref{ev}} is equal to some nonlocal gauge invariant operator at $\tau_0$, as discussed at the end of section~2.  Second, we can give it a path integral expression by inserting a `fold' at $\tau_0$ (Fig.~4a),
\begin{figure}[h]
\begin{center}
\vspace {-5pt}
\includegraphics[scale=1.2]{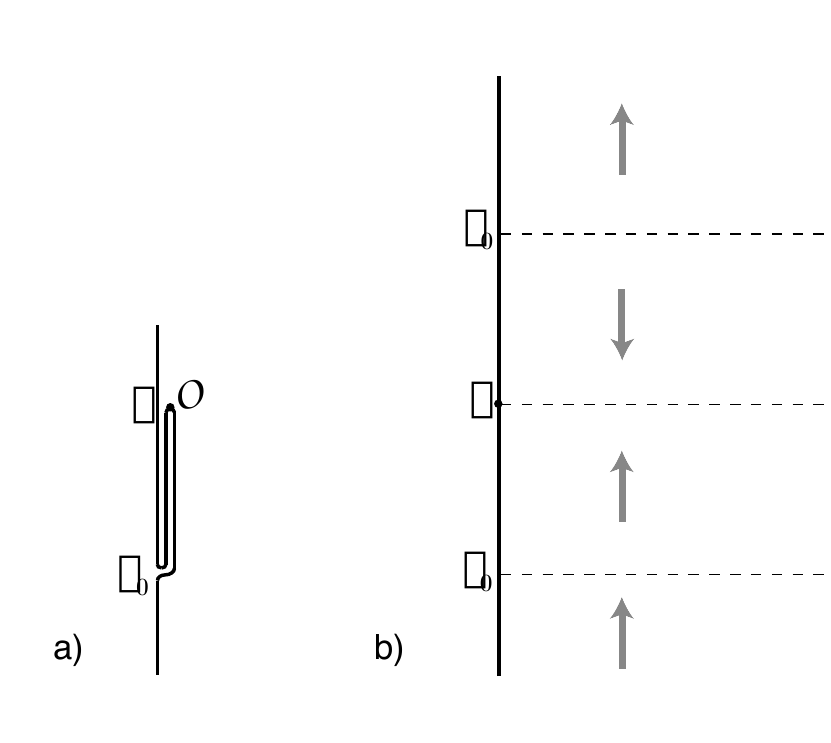}
\end{center}
\vspace {-10pt}
\caption{a) CFT path integral with a fold, to insert the operator ${\cal O}(\tau,x)$ at time $\tau_0$.  b) The fold unfolded, and extended into the bulk.  Arrows indicate the direction of time.}
\end{figure}
first integrating forward a time interval $\tau-\tau_0$, then inserting the operator ${\cal O}(x)$, and then integrating backwards in time by $-(\tau-\tau_0)$.  This is similiar to the Schwinger-Keldysh construction of real-time thermal Green's functions.  More generally, such a construction is needed whenever we wish to make a path-integral evaluation of correlators that are not time-ordered (though often this can be finessed by an analytic continuation of the time-ordered correlators).

This folded CFT should have a bulk dual, since we can think of the backwards evolution as a (finite-sized) perturbation of the CFT Hamiltonian,  changing it from $H$ to $-H$ in some intervals.  This can be obtained by a continuation through Euclidean values, $H \to e^{-i\theta} H$ with theta increased continuously from $0$ to $\pi$: each small change in $\theta$ corresponds to some CFT perturbation which is equivalent to some perturbation of the boundary conditions.  In fact, the $H$ of the CFT is the same as the bulk Hamiltonian $H_{\rm ADM}$, so in terms of some time-slicing this again corresponds to a folded path integral in the bulk, Fig.~4b.   See also the related discussion in \cite{Marolf:2008mg}.

We can now give a path integral version of the argument that the bulk-boundary map is independent of the Hamiltonian  at times other than $\tau_0$.  For convenience we take a coordinate system in which the Cauchy surface needed to define a gauge-invariant operator is the time slice $\tau = \tau_0$.  Now insert a backwards fold at $\tau_0 -\epsilon$ and a forward fold at $\tau_0+\epsilon$, Fig.~5a.
\begin{figure}[h]
\begin{center}
\includegraphics[scale=1.2]{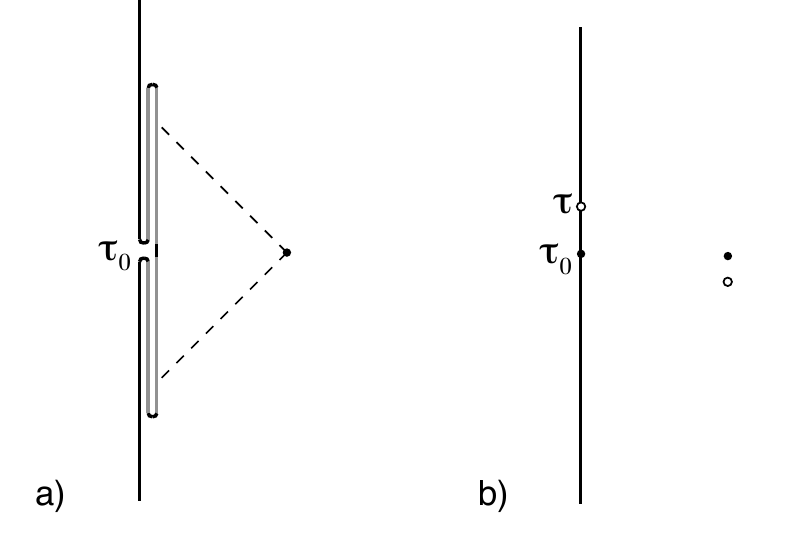}
\end{center}
\vspace {-10pt}
\caption{a) CFT path integral with a fold, to insert the operator ${\cal O}(\tau,x)$ at time $\tau_0$.  On the gray segments time evolution is generated by $H'$. b) Bulk fields ($\circ$, $\bullet$) and their corresponding boundary operators.  Insertion of $\circ$ appears to change the dynamics used to derive the map for $\bullet$.  However, each boundary operator should be understood in terms of a folded geometry as in (a), so that $\circ$ is in a different bulk region.}
\end{figure}
The two leaves of each fold cancel trivially, so we can take the Hamiltonian $H'$ on each leaf to be anything we wish.  Taking the folds long enough to capture the full smearing function for ${\cal O}(\tau_0, x)$, we can then integrate back using $H'$ (aside from the tiny interval $\epsilon > \tau > -\epsilon$) and then, when all operators are expressed in terms of the time $\tau_0$, the folds cancel and we are left with the original spacetime.  The map depends on the original CFT Hamiltonian $H$ only in an arbitrarily small neighborhood of $\tau_0$.   It is not clear whether this last dependence can be eliminated, as the constraints may relate bulk fields to sources at $\tau_0$. The map does not depend on $H'$, as we could have carried out the construction without introducing the fold at all.

The path integral construction clarifies the bulk interpretation of some measurements.  For example, Ref. \cite{Bousso:2012sj} argues that the bulk-to-boundary map does indeed depend on sources away from $\tau_0$.  A version of their argument is shown in Fig.~5b.  A boundary operator $\circ$ is introduced in such a way as to insert a bulk field $\circ$ with the bulk time-ordering opposite to that on the boundary.  Consider the bulk-boundary mapping for the $\bullet$ field/operator. The bulk $\circ$ appears to interfere with the back-evolution used in one derivation of the $\bullet$ map, and the boundary $\circ$  appears to interfere with the forward-evolution in the other derivation: the map is different than in the absence of the $\circ$ insertion.  The point is, however, that the two bulk operators in Fig.~5b lie on different folds of the bulk  (both of the sort shown in fig 5a): the bulk $\bullet$ on a fold at $\tau_0$, and the bulk $\circ$ on a fold at $\tau$, and so there is no interference.

Of course, one could certainly define a third CFT operator at another CFT time $\tau_1 > \tau$ that is also associated with the bulk field $\bullet$. The value of this new operator would then be affected by the insertion at $\circ$.  But this is just the usual story that quantum mechanical measurements at time $\tau_1$ can be affected by measurements at any time $\tau < \tau_1$.  In particular, even if both boundary times $\tau_0$ and $\tau_1$ are spacelike to $\bullet$, the two CFT operators associated to $\bullet$ can represent ``the same CFT operator'' only in the sense that, e.g., $x(t=0)$ and $x(t=2\pi/\omega)$ represent the same Heisenberg operator for a simple harmonic oscillator.  In this latter context, we immediately understand the sense in which measurements of $p(t=\pi/\omega)$ can affect $x(t=2\pi/\omega)$ but not $x(t=0)$.  The situation regarding the two CFT operators associated with $\bullet$ and the intervening operator $\circ$ is precisely the same.

\subsection{Other approaches}

\subsubsection{Holographic Renormalization Group}

A rather different way to think about the bulk fields is in terms of a change of variables in the path integral: we want to start with the path integral in the gauge theory and end up with the path integral over bulk string theory, or in the low energy approximation bulk gravitational field theory.  This requires that we in some way integrate the radial dimension of AdS into the system.  The holographic analog of the Wilsonian renormalization group~\cite{Heemskerk:2010hk,Faulkner:2010jy} suggests a framework for this.  Essentially we reverse-engineer the construction of Ref.~\cite{Heemskerk:2010hk}; the approach of Ref.~\cite{Lee:2010ub} is schematically similar.

Start with the CFT with a single-trace action at some cutoff scale~$\Lambda$.  Integrating the CFT fields down to a scale $\Lambda - \delta\Lambda$ generates a new action that contains double-trace terms~\cite{Li:2000ec}.  In order to restore the action to single-trace form, introduce an auxiliary field $\phi_{i,\Lambda}(x)$ for each single-trace operator ${\cal O}_i$, including tensor indices as needed.  Now iterate.  When the cutoff reaches $\Lambda = 0$, meaning that the CFT fields have been fully integrated out, what remains is the fields $\phi_{i,\Lambda}(x) \to \phi_{i}(\rho,x)$ where $\rho = 1/\Lambda$.

One thing not obvious in this approach is the locality of the effective action at scale
$\Lambda$.  As discussed in Ref.~\cite{Heemskerk:2009pn}, locality on the scale of the AdS radius $R$ is manifest, but not the locality expected on the shorter scales $l_{\rm string}$ and $l_{\rm Planck}$.  One might expect that, as in Ref.~\cite{Heemskerk:2009pn}, this could be derived (after a field redefinition) from the existence of the corresponding hierarchy in the CFT, but we have not completed the argument.  Further, the general covariance is not manifest.  Indeed, the holographic RG is simple only near geometries with a monotonic warp factor, so a more general framework is needed for arbitrary metrics.

\subsubsection{Two-point functions}

In our discussion, nonlocal gauge invariant operators can make measurements in the bulk of AdS spacetime.  Earlier work, beginning with Refs.~\cite{Balasubramanian:1999zv, Kraus:2002iv}, considered the bilocal product of local gauge invariant operators.  The expectation value of this receives contributions from spacelike geodesics and so seems to probe the interior.  However, this is is dual to a product of bulk fields near the boundary, and so should essentially commute with the fields deep in the interior.  Indeed, it was shown in Ref.~\cite{Louko:2000tp} that these observables cannot be used to make the kind of quantum measurement that we are discussing.

\subsubsection{Wilson loops}

Similar to the previous example, expectations values of spacelike Wilson loops involve spacelike world-sheets extending into the interior, and so seem to probe bulk physics~\cite{Susskind:1999ey}.  However, such operators create strings at the boundary, which should in the $1/N$ expansion commute with spacelike separated bulk fields.  Indeed, in parallel with the previous example, Refs.~\cite{Giddings:2001pt,Freivogel:2002ex} show that these cannot be used to make quantum measurements.

On the other hand, one expects that any gauge invariant operator can be written in terms of Wilson loops.  This is manifest with a lattice regulator, for example.  Thus the precursors that we measure should be expandable in such a basis.  How is this consistent with the conclusion above?  Refs.~\cite{Giddings:2001pt,Freivogel:2002ex} suggest that the precursors should be decorated Wilson loops, with insertions of local operators.  The CFT time evolution process that generates the precursors suggests a more extreme sore of decoration.  On the lattice it corresponds to attaching plaquettes randomly along the loop, a sort of branching diffusion.  The resulting paths are very highly kinked, much more so than a random walk, for example.  The continuum limit of such operators seems problematic, and so the expansion in Wilson loops may be only formal.  It would be good to make this more precise.

\subsubsection{Probes}

D-branes provide another means of probing the interior of a black hole~\cite{Horowitz:2009wm}.  Identifying their coordinates with the eigenvalues of the CFT scalars seems to give us what we asked for at the beginning of Sec.~4: a way to read bulk physics directly from the dynamics of the CFT, without having to also solve for bulk evolution.  Moreover, these observables refer more directly to the matrix degrees of freedom, which from various points of view are the origin of the emergent bulk.  Thus this seems like a promising direction to pursue.

However, things are not entirely so simple.\footnote{We thank M. Headrick and E. Silverstein for  discussions and suggestions on these points.}  First, as in any continuum quantum field theory, the true eigenvalues of the scalar fields are of order the UV cutoff, due to quantum fluctuations.  Also, the fluctuations of the commutators of different $\Phi^i$ are not parametrically smaller as compared to the $\Phi^i$ themselves (this is shown rigorously for the BFSS matrix theory in Ref.~\cite{Polchinski:1999br}), so there is not a well-defined eigenbasis.  To identify the dynamics of interest, one must first identify some effective low energy matrix fields.  A simple way to do this is to work with the chiral local operators, in the supersymmetric case.  At weak coupling these correspond directly to traces of eigenvalues, and their expectation values are UV-finite and (with enough supersymmetry) not renormalized at strong coupling.  Another approach, for extended probes such as D3-branes, is to use the eigenvalue of  the spatially averaged scalars, dealing with the issue of gauge invariance in some way.

Second, if some version of the eigenvalue is to be identified with the probe position, there is the question of time resolution.  To be precise, suppose we perturb the eigenvalue in the gauge theory: how and when does the probe in the bulk respond?  If we use the chiral operators to define the probe positions, the answer is clear: these are dual to the boundary values of supergravity fields, so one is effectively identifying the positions of probes with the long-ranged supergravity fields that they source on the moduli space.  But in the dynamical case, one must clearly wait a light-travel time for a boundary perturbation to be reflected in the probe's motion, and vice versa: this gives the minimum time resolution.  This may be adequate for some purposes, but it puts limits on the ability to distinguish the inside and outside of the horizon.  It is not clear that the spatially averaged eigenvalues noted above will behave similarly, but we expect that any observable that probes local physics in the bulk will be as complicated in the CFT as those we have considered.

\section*{Acknowledgments}
We thank Raphael Bousso, Steve Giddings, Matt Headrick, Veronika Hubeny, Dan Kabat, Jared Kaplan, Ami Katz, Albion Lawrence, Simon Ross, Eva Silverstein, and Lenny Susskind for useful discussions.
IH and JP were supported in part by NSF grants PHY07-57035 and PHY11-25915, and by FQXi grant RFP3-1017.  DM was supported in
part by NSF grants PHY08-55415 and PHY11-25915, by FQXi grant RFP3-1008,
and by funds from the University of California. He also thanks the Kavli Institute for Theoretical Physics
for their hospitality during much of this work.

\begin{appendix}

\sect{Spacelike Green's functions}

In this appendix we discuss the explicit construction of Green's functions for the wave equation in AdS whose support vanishes inside the past and future lightcones. The derivation of a spacelike Green's function was first demonstrated in \cite{Hamilton:2005ju,Hamilton:2006az} in even dimensions. Here we preface their construction by deriving an analogous spacelike Green's function in more familiar flat space, and extend the calculation to odd dimensions in both flat space and AdS.  Curiously, the odd-dimensional case is significantly more complicated: the spacelike Green's function cannot be taken to be a function of the invariant separation.  We also elucidate the light-cone behavior in the even-dimensional case, which is important for the causal properties of the Green's function. This affects some intermediate equations in Refs.~\cite{Hamilton:2005ju,Hamilton:2006az}, but not their final result for the smearing function.

\subsection{Minkowski space in even dimensions}

We follow here the approach of Refs. \cite{Hamilton:2005ju,Hamilton:2006az}, but applied to Minkoski spacetime.
The $d+1$ dimensional flat space wave equation, acting on a function of the invariant distance $\lambda = (x-x')^2$, reduces to 
\begin{equation}
4 \lambda G'' +  2 (d+1) G' -m^2 G = 0
\, .
\end{equation}
In Euclidean signature the general solution is
\begin{equation}
G_E(\sqrt{\lambda}) = \lambda^{-\mu/2}\left[ c_1 I_{\mu}(m\sqrt{\lambda}) + c_2 K_\mu(m\sqrt{\lambda}) \right], \qquad \mu = \frac{d-1}{2}\,.
\end{equation}
We consider the first case of integer $\mu$ (even-dimensional AdS). As $\lambda \to 0$ the first solution approaches a constant, while the leading singular behavior of the second term is given by 
\begin{equation}
\lambda^{-\mu/2} K_\mu(m\sqrt{\lambda}) \sim \frac{2^{\mu-1}\Gamma(\mu)}{m^{\mu}}\frac{1}{r^{d-1}} \, ,
\end{equation}
where $r=\sqrt{\lambda}$ is the Euclidean radial coordinate.
If we want $G_E(r)$ to be a normalized Euclidean Green's function $\Box G_E = \delta^{(d+1)}(x)$, we see that we should set
\begin{align}
c_1 = \textnormal{arbitrary}, \qquad c_2 = \frac{m^\mu}{(2\pi)^{\mu+1} }\, .
\end{align}
The Minkowski signature Feynman Green's function is related to the Euclidean Green's function by $G(\sqrt{\lambda}) = i G_E(\sqrt{\lambda+i\epsilon})$.

Next we will make use of the fact that the Bessel functions are related to the modified Bessel functions by
\begin{equation}
I_\mu(iz) = (- i)^{-\mu} J_\mu(z), \qquad
K_\mu(iz) = \frac{\pi}{2} i^{-\mu-1}  H^{(2)}_\mu(z),
\end{equation}
where the Hankel function of the second kind is defined as $H^{(2)}_\mu ( z) = J_\mu(z) - i Y_\mu(z)$. When we analytically continue to the timelike region, $\lambda<0$, we subsequently have
\begin{equation}
G_M(\sqrt{\lambda+i\epsilon})=G_M(i\sqrt{-\lambda}) = \frac{1}{ (-\lambda)^{\mu/2}}\left[ i c_1 J_\mu (m\sqrt{-\lambda}) + (-1)^\mu \frac{\pi c_2}{2} H^{(2)}_\mu(m\sqrt{-\lambda})\right]
\, .
\end{equation}
Since $G_M$ is a Green's function, so is the real part of $G_M$, which in the timelike region is
\begin{equation}
\Re G_M(i\sqrt{-\lambda})  = \frac{1}{ (-\lambda)^{\mu/2}}\left[ - \Im c_1 + (-1)^\mu \frac{\pi c_2}{2} \right]J_\mu(m\sqrt{-\lambda}) \qquad \textnormal{for } \lambda<0,
\end{equation}
and this vanishes if we set
\begin{equation}
\Im c_1 = (-1)^\mu \frac{\pi c_2}{2}
.
\end{equation}

The resulting Green's function {(which for brevity we refer to as spacelike, despite the fact that it is also non-vanishing at null separations)} is
\begin{equation}
\Re G_M(\sqrt{\lambda}) =  c_2 \left[ (-1)^{\mu+1} \frac{\pi}{2} \frac{I_{\mu}(m\sqrt{\lambda})}{\lambda^{\mu/2}}- \Im \frac{K_\mu(m\sqrt{\lambda+i\epsilon}) }{(\lambda+i\epsilon)^{\mu/2}} \right]
\, .
\end{equation}
For  $\lambda > 0$ the second term vanishes in the limit $\epsilon \to 0$ and the Green's function is given simply by the first term. The first term is exactly the homogeneous solution. It contributes no sources to the Green's function. The second term, however, does not vanish for $\lambda=0$. It is just the real part of the Feynman propagator and contributes a delta-function source at the origin. We thus see our solution has the correct source behavior to be a Green's function. It is also possible to see that the second term diverges on the light-cone, while the first term is regular. Thus we have the singular behavior required for a causal Green's function.

\subsection{Minkowski space in odd dimensions}

In odd dimensions the technique of the previous section again generates a solution of the equation of motion with spacelike and null support, but in this case it is a homogeneous solution, not a Green's function. Thus, there is no spacelike Green's function that is a function only of the invariant separation.  We have given in \S2 an argument that the Green's function should exist, so evidently it is not invariant under the AdS isometries.  This is possible because we are not dealing with a standard Cauchy problem, so the solution need not be unique.  A conformal transformation on the spacelike Green's function then gives a different spacelike Green's function.  The construction outlined in \S2 only guarantees invariance under $O(d-1)$ rotations.  We do not see a simple reason why the full conformal symmetry appears only in the even-dimensional case.

First consider the massless case. A general $O(d-1)$-invariant solution of the equation of motion takes the form
\begin{equation}
G(t,r) = \int_{-\infty}^{\infty}\frac{\d \omega}{2 \pi} e^{-i \omega t} \left( c_1(\omega) r^{1-n} J_{n-1}(|\omega| r  )+c_2(\omega) r^{1-n} Y_{n-1}(|\omega| r ) \right)
\, ,
\end{equation}
where $d=2n$. From Green's theorem applied to an infinitely long cylinder of finite radius around the origin, and using the property that our solution vanishes in the timelike region, we need to satisfy
\begin{equation}
\int_{S^{2n-1}}\d{\Omega} \int_{-\infty}^{\infty}\d{t} r^{2n-1} \partial_r G =1
\, .
\end{equation}
In the limit $r \rightarrow 0$, we then must have 
\begin{equation}
G = \frac{r^{2-2n}}{2(1-n)V_{S^{2n-1}}} \delta(t) + \sum_{i = 3-2n}^{0}r^ i P_i(\partial_t) \delta(t)
\, ,
\end{equation}
where $P_i$ are some finite degree polynomials. The first term in the preceding equation satisfies the Green's theorem requirement and the terms subleading in $1/r$ ensure that the Green's function has no timelike support as $r\rightarrow 0$. We have expansions of the Bessel functions near $r=0$ given by
\begin{align}
J_{n-1}(|\omega|r)&=  \left(\frac{|\omega|r}{2}\right)^{n-1}\frac{1}{\Gamma (n)} + O(r^n) \,, \\
Y_{n-1}(|\omega|r)&=  -\frac{1}{\pi}\left(\frac{|\omega|r}{2}\right)^{1-n}\sum_{k=0}^{n-2}\frac{\Gamma(n-1-k)}{\Gamma(k+1)}\left(\frac{|\omega|r}{2}\right)^{2k} + \frac{2}{\pi} \log\left(\frac{|\omega|r}{2}\right)\left(\frac{|\omega|r}{2}\right)^{n-1}\frac{1}{\Gamma(n)} \nonumber\\
& \qquad -\frac{1}{\pi}\left(\frac{|\omega|r}{2}\right)^{n-1}\frac{\psi(n)-\gamma}{\Gamma(n)}+ O(r^n)
\, .
\end{align}
Thus we conclude
\begin{align}
c_2(\omega) = \frac{ \pi}{2(n-1)V_{S^{2n-1}}\Gamma(n-1)}\left(\frac{|\omega|}{2}\right)^{n-1} = \frac{1}{4}\left(\frac{|\omega|}{2 \pi}\right)^{n-1}
\, .
\end{align}
All of the subleading terms from $Y_{n-1}$ that remain non-zero as $r\rightarrow0$ are now also polynomial in $\omega$, except for the logarithmic term. To eliminate this contribution, we simply set
\begin{equation}
c_1(\omega) = -\frac{2}{\pi} c_2 \log\left({|\omega|}\right) + C \left({|\omega|}\right)^{1-n}
\, ,
\end{equation}
where $C$ is an arbitrary constant. We thus have constructed the Green's function
\begin{align}
G(t,r) = \int_{-\infty}^{\infty}\frac{\d \omega}{2 \pi} e^{-i \omega t} \frac{1}{4}\left(\frac{|\omega|}{2 \pi}\right)^{n-1}\bigg[& \left(-\frac{2}{\pi} \log\left(|\omega|\right)+ C \left({|\omega|}\right)^{2-2n}\right) r^{1-n} J_{n-1}(|\omega| r  ) \nonumber\\
 &+ r^{1-n} Y_{n-1}(|\omega| r ) \bigg]
\, .
\end{align}

To extend this Green's function to the massive case, we need only replace $|\omega| \rightarrow \sqrt{\omega^2-m^2}$. 

\subsubsection{Position-space representation}

In the massless case only we are able to solve for these Green's functions explicitly in position space. Consider a solution of the form
\begin{equation}
G(\tau,r) = \frac{f(\tau)}{r^{2n-1}}+\frac{g(\tau)}{r^{2n-1}} \log(r)
\, ,
\end{equation}
where $\tau = t/r$. Plugging this ansatz into the equation of motion, we can explicitly find all such solutions and select the ones that vanish in the timelike region. One solution is the {imaginary} part of the standard Feynman solution. This has no source and is not a Green's function. The other solution is
\begin{equation}
G(\tau,r) = \Re\left[\frac{\log(r)+\log \left(\tau ^2-1\right)-(n-1) \tau ^2 \,
   _3F_2\left(1,1,1-n;\frac{3}{2},2;\tau ^2\right)}{r^{2n-1} \left(1-\tau ^2 + i\epsilon\right)^{(2n-1)/2}}\right]
\, .
\end{equation}
The hypergeometric function is just a finite degree polynomial in $\tau$ of degree $2n-{2}$ with real coefficients. By a quick inspection, it is clear that the real part of this solution {vanishes in the timelike region where $\tau >1$.}

It remains to check that this solution is a Green's function. We consider Green's theorem with a bounding surface given by an infinitely long cylinder of radius $r$. There is no contribution from the caps of the cylinder in the timelike region. Let us first consider the contribution of an arbitrary term in the polynomial along the side of the cylinder:
\begin{equation}
I_1 = \Re\left[\int_{-\infty}^{\infty} \d \tau \frac{\tau ^{2 n-2 m}}{ \left(1-\tau ^2+i \epsilon \right)^{\frac{(2n+1)}{2}}} \right]
\end{equation}
for $n>m>0$. The integrand falls sufficiently quickly so that we can rotate the contour to the imaginary axis, $\tau = i t$, and find the integral vanishes:
\begin{equation}
I_1 = \Re\left[i (-1)^{n-m} \int_{-\infty}^{\infty} \d t \frac{t ^{2 n-2 m}}{ \left(1+t^2 \right)^{\frac{(2n+1)}{2}}} \right] =0
\end{equation}
For $m=0$, however, in rotating the contour to the imaginary axis we need to keep the contribution from the connecting arc, which we take to be at fixed radius $|\tau|=R$ (and $R\rightarrow\infty$). As directly above, the contour along the imaginary axis gives no contribution to the real part. The arcs contribute
\begin{align}
I_2 &= \lim_{R\rightarrow\infty} \Re\left[ - \int_0^{\pi/2} \frac{i \d \theta (R e^{i \theta})^{2n+1}}{(1-R^2 e^{i 2\theta})^{n+1/2}} +\int_{\pi}^{3\pi/2} \frac{i \d \theta (R e^{i \theta})^{2n+1}}{(1-R^2 e^{i 2\theta})^{n+1/2}}   \right] \nonumber\\
 &= \lim_{R\rightarrow\infty} \Re\left[ \int_0^{\pi/2} \frac{i \d \theta}{(-1)^{n}(- i)} +\int_{\pi}^{3\pi/2} \frac{i \d \theta }{(-1)^{n}(-i)}   \right] \nonumber\\
 &= (-1)^{n} \pi
\end{align}
Lastly we compute the contribution from the integral
\begin{equation}
I_3 = \Re\left[\int_{-\infty}^{\infty} \d \tau \frac{\log(\tau^2-1-i \epsilon)}{ \left(1-\tau ^2+i \epsilon \right)^{\frac{(2n+1)}{2}}} \right]
\, .
\end{equation}
As in the first integral, we can rotate the contour to the imaginary axis to find
\begin{align}
I_3 &= \Re\left[i \int_{-\infty}^{\infty} \d t \frac{\log(-t^2-1)}{ \left(1+t^2\right)^{\frac{(2n+1)}{2}}} \right] \nonumber\\\
    &= - \pi \int_{-\infty}^{\infty} \d t \frac{1}{ \left(1+t^2\right)^{\frac{(2n+1)}{2}}} \nonumber \\
    &= -\pi^{3/2} \Gamma(n)/\Gamma(n+1/2)
\, .
\end{align}

We see that the integrals are all finite. Thus we are able to normalize our solution such that this is a Green's function with a simple delta function source. 

\subsection{Anti-de Sitter in even dimensions}

We begin with the wave equation in AdS$_{d+1}$ ($d+1=2n$) acting on a function of the invariant distance
\begin{equation}
\sigma = X\cdot X' = R^2 \frac{\cos(\tau-\tau')-\sin\rho\sin\rho'\Omega\cdot\Omega'}{\cos\rho\cos\rho'} ,
\end{equation}
which reduces to
\begin{equation} \label{AdSInvariantWaveEqn}
(\sigma^2-1) G'' + (d+1)\sigma G' - \Delta(\Delta-d) G  =0
\, .
\end{equation}
We consider this equation in the Euclidean regime where $\sigma > 1$. The general solution is 
\begin{equation} \label{generalsoln}
G_E(\sigma) = (\sigma^2-1)^{-\mu/2} \left[ c_1 \mathbf{P}^\mu_\nu(\sigma) + c_2 \mathbf{Q}^\mu_\nu (\sigma) \right]
\, .
\end{equation}
Here $\mathbf{P}, \mathbf{Q}$ are the Legendre functions of type 3, which have branch cuts from $-\infty$ to $-1$ and $-\infty$ to 1. We have also defined $\mu = (d-1)/2$ and $\nu = \Delta - (d+1)/2$. We consider the case $\mu\in \N$.

Following closely \cite{Hamilton:2005ju,Hamilton:2006az}, we derive a spacelike AdS Green's functions in parallel to the flat-space case. The $\mathbf{Q}$ solution has the right short distance behavior to be the Euclidean Green's function, and demanding that its source is a normalized delta function fixes the coefficient $c_2$. We again have the freedom to add a homogeneous solution, leaving $c_1$ undetermined. 

Continuing the general solution \eqref{generalsoln} into the lightcone $\sigma<1$, we make use of the identity
\begin{equation}
(z^2-1)^{-\mu/2} \mathbf{Q}_{\nu }^\mu(z)=(-1)^\mu (1-z^2)^{-\mu/2} \left(Q_{\nu
   }^\mu(z)-i \frac{\pi }{2}P_{\nu }^\mu(z)\right)
   \, ,
\end{equation}
where ${P}, {Q}$ are the Legendre functions of type 2. This should be seen as exactly analogous to the flat-space identity above. 

Defining $G_M(\sigma) = i G_E(\sigma+i\epsilon)$ one obtains, in the time-like region 
\begin{align} \label{shortdistancecontinuation}
G_M(\sigma) = i c_1 (-1)^\mu (1-\sigma^2)^{-\mu/2}P_{\nu }^\mu(\sigma) + ic_2 (-1)^\mu (1-\sigma^2)^{-\mu/2} \left[Q_{\nu
   }^\mu(\sigma)-i \frac{\pi }{2}P_{\nu }^\mu(\sigma) \right]
\end{align}
We see that the imaginary part of $c_1$ can be chosen to cancel the $P_{\nu }^\mu$ term inside the lightcone if we take $c_1 = i \pi c_2 /2$. 

Again, $\Re G_M$ is still a Green's function. The above choice of $c_1= i \pi c_2/2$ makes $\mathrm{Re}G_M$ vanish inside the lightcone; it is only non-timelike supported.
\begin{equation}
\Re G_M = -c_2 \left[ \frac{\pi}{2} (\sigma^2-1)^{\mu/2}\mathbf{P}_{\nu }^\mu(\sigma) + \Im ((\sigma+i\epsilon)^2-1)^{-\mu/2} \mathbf{Q}_{\nu}^\mu(\sigma+i\epsilon)\right]
\end{equation}
This Green's function has the correct singular behavior. The first term is a homogeneous solution to the wave equation and smooth. It contributes no singularities on the lightcone nor sources. The second term is the Feynman propagator in AdS. It contributes a source at the origin and is a distribution supported on the light-cone.  In Ref.~\cite{Hamilton:2006az}, only the first, fully spacelike, term was retained in taking the real part.  This gives the correct smearing function due to the falloff of $ \mathbf{Q}_{\nu}^\mu$ at the boundary; however, the full Green's function would be needed in order to treat interactions.

\subsection{AdS in odd dimensions}

As in flat space, we consider solutions of the form $e^{-i \omega t}G(\rho)$. The equation of motion becomes:
\begin{equation}
 \omega ^2 \cos ^2(\rho ) G(\rho ) +\frac{\cos ^{d+1}(\rho )}{\sin ^{d-1}(\rho )} \partial_\rho \left(\frac{\sin ^{d-1}(\rho )}{\cos ^{d-1}(\rho )}G'(\rho )\right)-\Delta  (\Delta -d) G(\rho )=0 
\, .
\end{equation}
A general solution is given by
\begin{align}
G(\rho,t) = \int \frac{\d \omega}{2 \pi} e^{-i \omega t}\bigg[& c_1(\omega) G_1(\rho,\omega)+c_2(\omega) G_2(\rho,\omega)\bigg]
   \, 
\end{align}
with
\begin{align}
G_1(\rho,\omega) &=  \cos ^{d-\Delta }(\rho ) \, _2F_1\left(\frac{1}{2} (d-\Delta -\omega),\frac{1}{2} (d-\Delta +\omega );\frac{1}{2} (d-2 \Delta +2);\cos ^2(\rho )\right)\,, \nonumber\\
G_2(\rho,\omega) &=  \frac{ \Gamma \left(\frac{\Delta }{2}-\frac{\omega }{2}\right) \Gamma \left(\frac{\Delta }{2}+\frac{\omega }{2}\right) }{\Gamma (n) \Gamma (\Delta -n)}\left(1-\sin ^2(\rho )\right)^{\Delta /2} \, _2F_1\left(\frac{\Delta }{2}-\frac{\omega }{2},\frac{\Delta }{2}+\frac{\omega }{2};n;\sin ^2(\rho )\right)
\, ,
\end{align}
where again we have taken $d=2n$. In odd dimensions, the first solution is written in terms of a confluent hypergeometric function. For $\rho \ll 1$, it is more useful to rewrite this  function in the form
\begin{multline}
\, _2F_1\left(n-\frac{\Delta }{2}-\frac{\omega }{2},n-\frac{\Delta }{2}+\frac{\omega }{2};n-\Delta +1;\cos ^2(\rho
   )\right)= \\
   \frac{(-1)^{-n} \Gamma (n-\Delta +1) \log \left(\sin ^2(\rho )\right) \, _2F_1\left(n-\frac{\Delta
   }{2}-\frac{\omega }{2},n-\frac{\Delta }{2}+\frac{\omega }{2};n;\sin ^2(\rho )\right)}{(n-1)! \Gamma \left(-\frac{\Delta
   }{2}-\frac{\omega }{2}+1\right) \Gamma \left(-\frac{\Delta }{2}+\frac{\omega }{2}+1\right)} \\
   +\sum_{k=0}^{n-2}\frac{(n-2)! \Gamma (n-\Delta+1)  \left(-\frac{\Delta }{2}-\frac{\omega }{2}+1\right)_k \left(-\frac{\Delta}{2}+\frac{\omega }{2}+1\right)_k \left(\sin ^2(\rho )\right)^{k-n+1}}{k! (2-n)_k \Gamma \left(n-\frac{\Delta }{2}-\frac{\omega }{2}\right) \Gamma
   \left(n-\frac{\Delta }{2}+\frac{\omega }{2}\right)} \\
   +\sum_{k=0}^{\infty}\frac{(-1)^{1-n}  \Gamma (n-\Delta +1)
   \left(n-\frac{\Delta }{2}-\frac{\omega }{2}\right)_k \left(n-\frac{\Delta }{2}+\frac{\omega }{2}\right)_k \left(\sin ^2(\rho )\right)^k}{k! (k+n-1)! \Gamma \left(-\frac{\Delta }{2}-\frac{\omega }{2}+1\right)
   \Gamma \left(-\frac{\Delta }{2}+\frac{\omega }{2}+1\right)} \bigg(-\psi
   ^{(0)}\left(k+n-\frac{\Delta }{2} -\frac{\omega }{2}\right)\\
   -\psi ^{(0)}\left(k+n-\frac{\Delta }{2}+\frac{\omega
   }{2}\right)+\psi ^{(0)}(k+n)+\psi ^{(0)}(k+1)\bigg)
\, .
\end{multline}

Near $\rho=0$ the AdS metric approaches the flat space metric $\d s^2 = \d t^2+\d \rho^2 + \rho^2 \d \omega_{d-1}^2$. Hence, we proceed exactly as in the flat space case in the preceding section. We fix the coefficient of the leading $\rho^{-2n+2}$ term to give the correct surface integral for small $\rho$. We want a solution of the form
\begin{equation}
G(\rho,t) = \frac{1}{(2-2n)V_{S^{2n-1}}}\rho^{2-2n} \delta(t) + \sum_{i=3-2n}^0 P_i(\partial_t)\delta(t)\rho^i + O\left(\rho\right) \,,
\end{equation}
where $V_{S^{2n-1}}=2 \pi^{n}/\Gamma(n)$. Fixing the leading term constrains
\begin{equation}
c_1 = -\frac{\Gamma \left(n-\frac{\Delta }{2}-\frac{\omega }{2}\right) \Gamma
   \left( n-\frac{\Delta}{2} +\frac{\omega}{2} )\right)}{4 \pi^n \Gamma (n-\Delta +1)}
\, .
\end{equation}
All of the subleading terms in $c_1 G_1$ contain only finite, positive powers of $\omega$, and hence still have support only at $t=0$, until we reach $O(\rho^0)$. We need to cancel the contribution from the polygamma functions in $G_1(\rho)$ at order $O(\rho^0)$, which have support away from $t=0$. This cancellation determines $c_2$ to be
\begin{equation}
c_2 = -c_1 \frac{(-1)^n \Gamma (n-\Delta +1) \Gamma (\Delta -n) \left(H_{n-\frac{\Delta }{2}-\frac{\omega }{2}-1}+H_{n-\frac{\Delta }{2}+\frac{\omega }{2}-1}\right)}{\Gamma
   \left(\frac{\Delta -\omega }{2}\right)\Gamma
   \left(\frac{\Delta +\omega }{2}\right) \Gamma \left(1-\frac{\Delta }{2}-\frac{\omega }{2}\right) \Gamma \left(1-\frac{\Delta }{2}+\frac{\omega }{2}\right) }
\, ,
\end{equation}
where $H_x$ is the Harmonic Number. We thus have the solution
\begin{align}
G(\rho,\omega) =& -\frac{\Gamma \left(n-\frac{\Delta }{2}-\frac{\omega }{2}\right) \Gamma \left(n-\frac{\Delta }{2}+\frac{\omega }{2}\right)}{4 \pi ^{n}\Gamma (n-\Delta +1)}\bigg( G_1(\rho) \\
&-\frac{(-1)^n \Gamma (n-\Delta +1) \Gamma (\Delta -n) \left(H_{n-\frac{\Delta }{2}-\frac{\omega }{2}-1}+H_{n-\frac{\Delta }{2}+\frac{\omega }{2}-1}\right)}{\Gamma
   \left(\frac{\Delta -\omega }{2}\right)\Gamma
   \left(\frac{\Delta +\omega }{2}\right) \Gamma \left(1-\frac{\Delta }{2}-\frac{\omega }{2}\right) \Gamma \left(1-\frac{\Delta }{2}+\frac{\omega }{2}\right) } G_2(\rho)\bigg)
\end{align}
For small $\tilde \rho= \pi/2- \rho$, the coefficient of the term proportional to $\tilde\rho^{2n-\Delta}$ is
\begin{equation}
-\frac{\Gamma \left(n-\frac{\Delta }{2}-\frac{\omega }{2}\right) \Gamma \left(n-\frac{\Delta }{2}+\frac{\omega }{2}\right)}{4 \pi ^{n}\Gamma (n-\Delta +1)}\left(1-\frac{(-1)^n \Gamma (n-\Delta +1) \Gamma (\Delta -n) \left(H_{n-\frac{\Delta }{2}-\frac{\omega }{2}-1}+H_{n-\frac{\Delta }{2}+\frac{\omega }{2}-1}\right)}{\Gamma
   \left(\frac{\Delta -\omega }{2}\right)\Gamma
   \left(\frac{\Delta +\omega }{2}\right) \Gamma \left(1-\frac{\Delta }{2}-\frac{\omega }{2}\right) \Gamma \left(1-\frac{\Delta }{2}+\frac{\omega }{2}\right) }\right)
\end{equation}
After some algebra, and using Green's theorem, this determines a tentative smearing function
\begin{equation}
k(\omega) = \frac{(-1)^{n} \Gamma (\Delta -n+1)}{2 \pi ^{n}\Gamma
   \left(\frac{\Delta}{2}-n+1 +\frac{\omega}{2}\right)\Gamma
   \left(\frac{\Delta}{2}-n+1 -\frac{\omega}{2}\right)} \left(H_{\frac{\Delta }{2}-n -\frac{\omega }{2}-1}+H_{\frac{\Delta }{2}-n+\frac{\omega }{2}} \right)
\end{equation}
Let us now compare this to the Fourier transform of the previously found \cite{Hamilton:2005ju,Hamilton:2006az} smearing function in odd dimensions, K(t), given by
\begin{equation}
K(t) = C (\cos(t))^{\Delta-2n}\log(\cos(t))\theta(t-\pi/2)\theta(t+\pi/2)
\end{equation}
We find the Fourier transform to be
\begin{align}
& C \int_{-\pi/2}^{\pi/2}\d \omega  e^{-i \omega t } (\cos(t))^{\Delta-2n}\log(\cos(t))=\\
&-C \frac{\pi  2^{2 n-\Delta -1} \Gamma (-2 n+\Delta +1) \left(H_{\frac{1}{2} (-2 n+\Delta -\omega )}+H_{\frac{1}{2} (-2
   n+\Delta +\omega )}-2 H_{\Delta -2 n}+\log (4)\right)}{\Gamma
   \left(\frac{\Delta}{2}-n+1 +\frac{\omega}{2}\right)\Gamma
   \left(\frac{\Delta}{2}-n+1 -\frac{\omega}{2}\right)}
\end{align}
It can be seen immediately that the first two Harmonic Number terms of the Fourier transform of the smearing function match to our solution by fixing
\begin{equation}
C = \frac{(-1)^{n-1}2^{\Delta-2n} \Gamma (\Delta-n +1)}{\pi ^{n+1}\Gamma (-2 n+\Delta +1)} \,.
\end{equation}
This now exactly matches the smearing function as written in \cite{Hamilton:2005ju,Hamilton:2006az}, except for the term
\begin{equation}
C \frac{\pi  2^{2 n-\Delta-1 } \Gamma (-2 n+\Delta +1) \left(-2H_{\Delta -2 n}+\log (4)\right)}{\Gamma
   \left(\frac{\Delta}{2}-n+1 +\frac{\omega}{2}\right)\Gamma
   \left(\frac{\Delta}{2}-n+1 -\frac{\omega}{2}\right)} \,.
   \end{equation}
We can match this extra piece simply by the addition of
\begin{equation}
-\frac{\left(-\frac{1}{\pi }\right)^n \Gamma (-n+\Delta +1) \left(H_{\Delta -2 n}-\log (2)\right)}{ \Gamma
   \left(\frac{\Delta}{2}-n+1 +\frac{\omega}{2}\right)\Gamma
   \left(\frac{\Delta}{2}-n+1 -\frac{\omega}{2}\right)} G_2(\rho) \,.
\end{equation}
to our solution. Since $G_2$ goes as $O(\rho^0)$, this adds no new singular terms for small $\rho$. Moreover, the denominator above is such that the $O(\rho^0)$ term is polynomial in  $\omega$ and so has delta-function support. It is just a homogeneous solution, which we are free to add. 

Thus we have reproduced the smearing function found in \cite{Hamilton:2005ju,Hamilton:2006az} in odd dimensions from a spacelike supported Green's function. 

\end{appendix}

\end{document}